\documentclass[preprint,twocolumn,5p,authoryear,compress,colorlinks=true]{elsarticle}

\journal{Icarus}

\usepackage{lineno,hyperref,siunitx,mhchem,booktabs,cuted,titlesec,afterpage}
\modulolinenumbers[1]

\usepackage[labelfont=bf,justification=justified,singlelinecheck=false]{caption}
\makeatletter
\g@addto@macro\@floatboxreset\centering
\makeatother

\newcommand{\DOI}[1]{%
 \newline\noindent\small%
 \textit{\small{DOI:} }%
 \href{https://dx.doi.org/#1%
 }{#1%
 }%
}
\newcommand{\figref}[1]{\autoref{fig:#1}}
\newcommand{\tabref}[1]{\autoref{tab:#1}}

\begin{document}

\begin{frontmatter}
\title{Mapping polar atmospheric features on Titan with VIMS:\\from the dissipation of the northern cloud\\to the onset of a southern polar vortex}

\author[LPG]{St\'ephane {Le Mou\'elic}\corref{correspondingauthor}}\ead{stephane.lemouelic@univ-nantes.fr }
\author[IPGP]{S\'ebastien Rodriguez}
\author[LPG]{Rozen Robidel}
\author[IPAG]{Baptiste Rousseau}
\author[LPG]{Beno\^{i}t Seignovert}
\author[JPL]{~\\Christophe Sotin}
\author[DPUI]{Jason W. Barnes}
\author[DPS]{Robert H. Brown}
\author[JPL]{Kevin H. Baines}
\author[JPL]{Bonnie J. Buratti}
\author[PSI]{Roger N. Clark}
\author[Cornell]{~\\Philip D. Nicholson}
\author[GSMA]{Pascal Rannou}
\author[ESAC]{Thomas Cornet}

\address[LPG]{LPG, UMR 6112, CNRS, Universit\'e de Nantes, 2 rue de la Houssini\`ere, 44322 Nantes, France}
\address[IPGP]{IPGP, CNRS-UMR 7154, Universit\'e Paris-Diderot, USPC, Paris, France}
\address[IPAG]{IPAG, CNRS, Universit\'e Grenoble Alpes, 38000 Grenoble, France}
\address[JPL]{Jet Propulsion Laboratory, California Institute of Technology, Pasadena, CA 91109, USA}
\address[DPUI]{Department of Physics, University of Idaho, Engineering-Physics Building, Moscow, ID 83844, USA}
\address[DPS]{Department of Planetary Sciences, University of Arizona, Tucson, AZ 85721, USA}
\address[PSI]{Planetary Science Institute, Tucson, USA}
\address[Cornell]{Department of Astronomy, Cornell University, Ithaca, NY 14853, USA}
\address[GSMA]{GSMA, Universit\'{e} de Reims Champagne-Ardenne, UMR 7331-GSMA, 51687 Reims, France}
\address[ESAC]{European Space Astronomy Centre (ESA/ESAC), Villanueva de la Canada, Madrid, Spain}

\cortext[correspondingauthor]{Corresponding author}

\begin{abstract}
We have analyzed the complete archive of the Visual and Infrared Mapping Spectrometer (VIMS) data in order to monitor and analyze the evolution of the clouds and haze coverage at both poles of Titan during the entire Cassini mission. Our objective is to give a cartographic synopsis from a VIMS perspective, to provide a global view of the seasonal evolution of Titan's atmosphere over the poles. We leave the detailed comparison with the Imaging Science Subsystem (ISS) and the Composite Infrared Spectrometer (CIRS) data sets to further studies. We have computed global hyperspectral mosaics for each of the 127 targeted flybys of Titan to produce synthetic color maps emphasizing the main atmospheric features.
The north pole appears fully covered by a huge cloud as soon as the first observations in 2004 and up to the equinox in 2009 \citep{LeMouelic2012}. The northern skies then became progressively clearer, after the circulation turnover in 2009, revealing the underlying lakes and seas to the optical instruments up to 2017. The reverse situation is observed over the south pole, which was mostly clear of such a high obscuring cloud during the first years of the mission, but started to develop a polar cloud in 2012. This feature grew up month after month until the end of the mission in 2017, with a poleward latitudinal extent of \ang{75}S in 2013 up to \ang{58}S in April 2017.
Thanks to the spectral capabilities of VIMS, we have detected HCN spectral signatures over the north pole in almost all flybys between 2004 and 2008. These HCN signatures started then to show up over the south pole in almost all flybys between 2012 and 2017, so perfectly matching the timing and spatial extent of the northern and southern polar atmospheric features.
\end{abstract}

\begin{keyword}
Titan \sep Titan atmosphere \sep Titan clouds \sep Image processing \sep Infrared observations
\DOI{10.1038/s41550-018-0434-z}
\end{keyword}

\end{frontmatter}


\section{Introduction}
The Cassini spacecraft entered in Saturn's orbit in July 2004. In thirteen years of operations, 127 targeted flybys of Titan, the biggest satellite of Saturn, have been performed. Titan is one of the most intriguing planetary bodies in the solar system, with a dense nitrogenous atmosphere and thick layers of atmospheric aerosols which completely mask the surface at visible wavelengths. We focus our study on the analysis of the global Cassini/VIMS (Visual and Infrared Mapping Spectrometer) data set acquired between 2004 and 2017, with a particular emphasis on the atmospheric polar features.
The objective is to document the seasonal behavior of photochemical haze and the formation and evolution of clouds in the winter polar vortex. Convective methane clouds have generally been observed at \ang{40}S and at the pole in the summer hemisphere, in addition to a global subsiding ethane cloud in the winter hemisphere \citep{Griffith2006,Hirtzig2006,Rannou2006,Rodriguez2011}. First evidences for a vast ethane cloud covering the north pole have been reported as early as the second targeted flyby of Titan in December 2004 \citep{Griffith2006}.
The first detailed imaging of this north polar feature with VIMS was obtained in December 2006, thanks to a change in inclination of the spacecraft orbit \citep{LeMouelic2012}.
During the first years of the mission, the northern lakes and seas that were not in winter night, were totally masked to the optical instruments by thick haze and clouds, whereas the surface of the southern polar regions was well illuminated and clearly visible in the methane windows \citep{Porco2005,Barnes2009,Turtle2009}.
Subsequent flybys revealed that the north polar atmospheric feature was progressively vanishing around the equinox in 2009 \citep{Rodriguez2009,Rodriguez2011,Brown2010,LeMouelic2012}, consistently with the results of Global Circulation Models \citep{Rannou2006}.
It revealed progressively the underlying lakes to the ISS (Imaging Science Subsystem) and VIMS instruments \citep{Turtle2011,Barnes2011}. Over the south pole, a high altitude cloud appeared in VIMS and ISS images acquired in May 2012 \citep{deKok2014,West2016}, within which condensates of HCN were identified \citep{deKok2014,Teanby2017}.

In this study, we have computed individual global maps of the north and south poles for each targeted flyby of Titan, using VIMS wavelengths sensitive both to clouds and surface features. This allows a more complete and detailed monitoring of the evolution of the north polar cloud than was done before by using a selection of individual flybys only \citep{LeMouelic2012}.
This study largely extends the time period of the monitoring, from 2004 to 2017. It also provides a detailed investigation of what has been acting over the south pole from the beginning up to the end of the Cassini mission. In the first section, we describe the data processing that we have performed on the global VIMS archive. The second section focuses on the evolution of the north polar features, from the first observations of the vast global cloud to its complete dissipation. The third section presents the concomitant evolution over the south pole. The fourth section discusses the spectral detection of the HCN at both poles, before concluding.

\section{Data processing}

During 13 years of operation, VIMS has acquired more than \num{60000} hyperspectral data cubes of Titan. The general description of the VIMS instrument can be found in \cite{Brown2004}. The VIMS IR part of the instrument acquires individual images up to $64 \times 64$ pixels, with a \SI{256}{wavelengths} spectrum between \num{0.88} and \SI{5.10}{{\um}} for each pixel. We applied the standard radiometric calibration pipeline described in \cite{Barnes2007}, and further refined by \cite{Clark2018}, labeled RC19. Within all the VIMS observations of Titan, we have systematically analyzed all the cubes covering the two polar regions taken between the T0 (2 July 2004) and T126 (17 May 2017) flybys. The surface of Titan can be observed through seven infrared methane windows at \num{1.08}, \num{1.27}, \num{1.59}, \num{2.01}, \num{2.69}, \num{2.78} and \SI{5}{{\um}} \citep{Sotin2005}. However, it should be pointed out that the surface feature observations contain portions of the atmosphere even within these methane windows due to opacity.

\begin{figure}[!ht]
 \includegraphics[width=.8\linewidth]{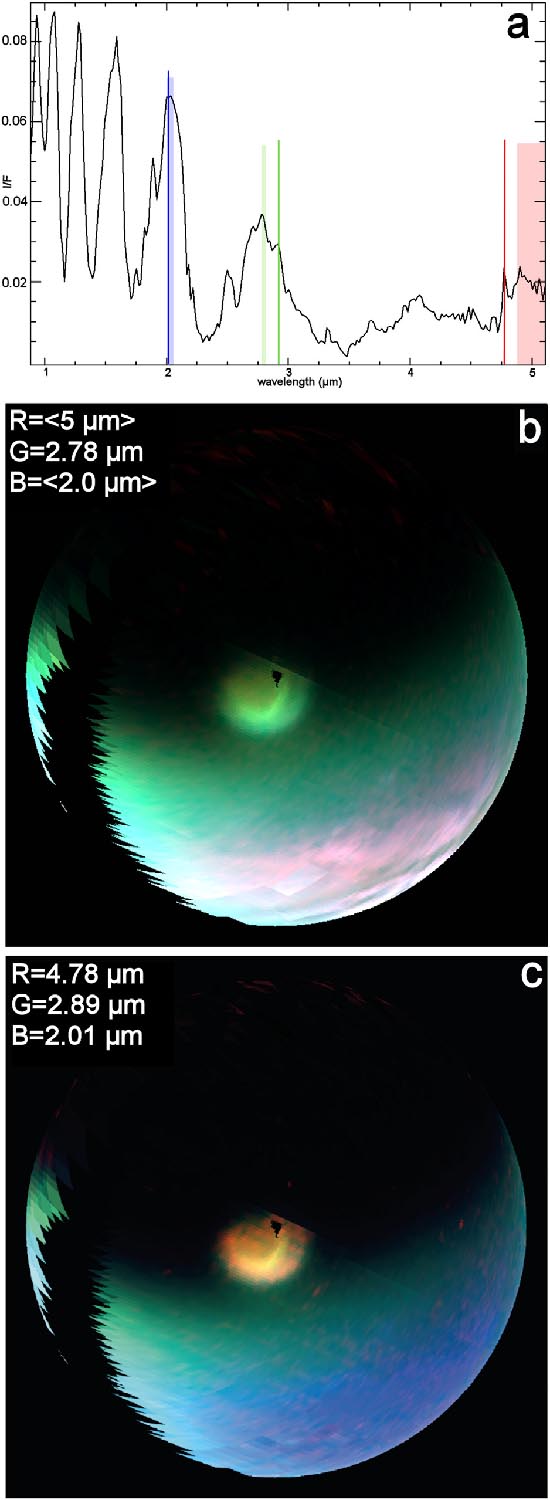}
 \caption{(a) Typical VIMS spectrum taken in northern latitudes, with colored bars indicating the wavelengths used for the RGB composites. (b) and (c): examples of an RGB mosaic of all VIMS data acquired during the T91 flyby (23 May 2013), using an orthographic projection centered on the south pole. The second RGB composition (c) is specifically designed to identify the condensed HCN cloud extension (see text).}
 \label{fig:spectra}
\end{figure}

\clearpage

In a first step, we have computed for each Titan flyby a global mosaic of all the individual VIMS cubes, after sorting the cubes by increasing spatial resolution. A visual inspection has been carried out to remove spurious individual images affected by calibration issues or too extreme observing geometries. These mosaics have then been projected using an orthographic projection centered on the north and south poles.
It should be noted that when the viewing direction is not straight over the pole, the clouds at different altitudes might appear slightly shifted compared to the surface. In order to extract the relevant information from this set of hyperspectral mosaics, we computed two series of RGB false color composites, which have been designed to catch both atmospheric and surface features.
The dynamic of the color stretch is the same for all the images throughout the mission, so that all the maps can be compared with each other. Our first color composite is similar to the one used in the \cite{LeMouelic2012} study of the north polar region, with the red corresponding to the average of 14 spectral channels between \num{4.88} and \SI{5.10}{\um}, the green to the \SI{2.78}{\um} channel and the blue to the average of five spectral channels between \num{2.00} and \SI{2.06}{\um} (pink, light green and light blue bars in the spectrum of \figref{spectra}a and corresponding image in \figref{spectra}b). The northern and southern polar atmospheric data also display a particularly bright and diagnostic spectral signal at \SI{4.78}{\um} related to the presence of condensed (solid or liquid) HCN \citep{Clark2010,deKok2014}.
We therefore computed a second set of RGB composites with the red controlled by the \SI{4.78}{\um} channel, the green by the \SI{2.89}{\um} channel and the blue by the \SI{2.01}{\um} channel (red, green and blue bars in the spectrum of \figref{spectra}a and corresponding image in \figref{spectra}c). This second set of maps is designed to catch the presence of the condensed HCN component, which would appear with a dominance of reddish tone, and its evolution with time. This will be discussed latter in the paper.

\section{Temporal evolution of the north polar atmospheric features during 13 years: from a \emph{mammoth cloud} to clear skies}

\figref{npole_all} shows the complete set of flybys' mosaics orthographically re-projected over the north pole, using the second RGB composite emphasizing the \SI{4.78}{\um} spectral feature. Observations were acquired at varying observing geometries and spatial resolutions, depending on both orbital constraints and instrument time allocation during each flyby. This figure illustrates the time evolution of the observed features, independently of the illumination/acquisition geometries and the quality of the available data. We decided to use the same color stretch, with the same thresholds, to display all the images, so that they can be directly compared with each other. This complete set of re-projected mosaics allowed us to select the images representative of the main evolution stages of the clouds structures.

\figref{npole_cloud} displays a selection of representative mosaics over the north pole from T0 to T126. We choose to display these mosaics using the color composite with red = \SI{5}{\um}, green = \SI{2.78}{\um} and blue = \SI{2.0}{\um}, showing more finely the seasonal evolution of the polar haze cap opacity and enhancing the surface features when distinguishable. This northern area was partly in the polar night during the northern winter, from October 2002 to August 2009.
The north polar cloud is detected very early in the mission \citep{Griffith2006}, as shown also on our Tb mosaic (December 2004) in the upper left view in \figref{npole_cloud}. One can distinguish in greenish tones the outer boundary limit of the most opaque part of the haze which extends up to about \ang{45}-\ang{50}N \citep{Rannou2012}. 
The core of the cloudy structure begins poleward of \ang{70}N (distinguishable in orange), but is not entirely visible due to the polar night. The same feature can be seen in the T10 mosaic (15 January 2006), indicating that the structure of the cloud is only slowly varying in this period of time. 
The new observation geometry at T23 offers a vision of the polar cloud with a better contrast, thanks to a change in inclination of Cassini's orbit and lower emergence angles. 
For the first time, we can distinguish subtle longitudinal heterogeneities within the north polar cloud.

\afterpage{%
    \begin{figure*}[!ht]
    \includegraphics[width=.95\linewidth]{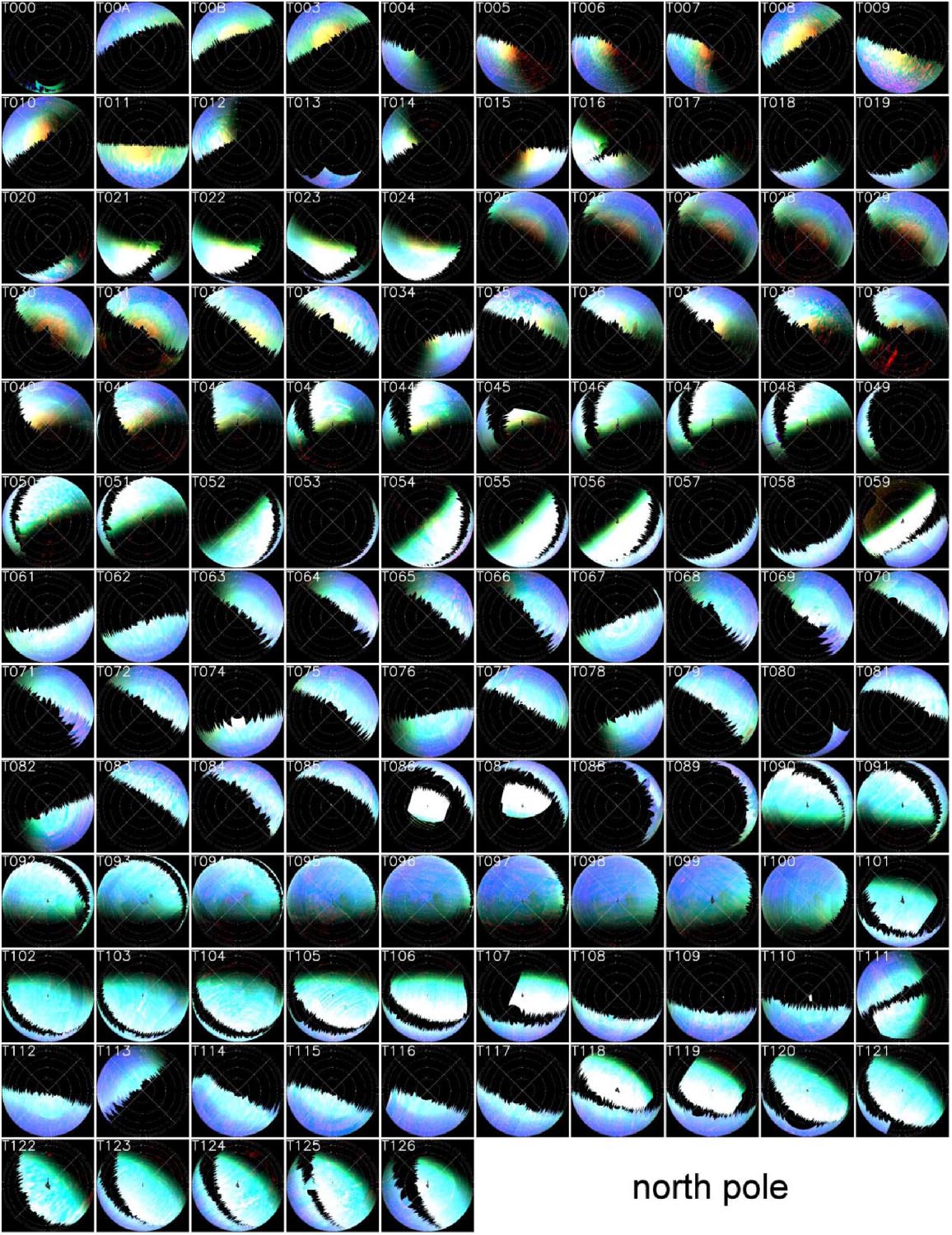}
    \caption{Systematic mapping of the north pole using all targeted Titan flybys from T0 (July 2004) to T126 (April 2017). Colors have been coded using the \SI{4.78}{\um}, \SI{2.89}{\um} and \SI{2.01}{\um} channels for the red, green and blue, respectively.}
    \label{fig:npole_all}
    \end{figure*}
\clearpage
}

A slight thinning of the cloudy and hazy structure is observed near the north pole between T38 (5 December 2007) and T46 (3 November 2008). The most obvious start of the cloud breakdown is emphasized in the T51 flyby of March 27, 2009, due to favorable observing conditions. The inner part of the cloud clearly starts to dissipate and reveals the underlying lakes, leaving only an outer ring. This structure continues to vanish progressively, and in June 2010 (T69, almost a year after the equinox), only a small residual can be detected right above the illuminated north pole. This confirms the trend that was suspected in \cite{LeMouelic2012} using an analysis of flybys up to T56 only.
Then, as shown in April 2013 (T90), September 2013 (T94), April 2014 (T100) and August 2014 (T104), VIMS started to acquire a series of its best mosaics of the sunlit north polar region, where surface features such as lakes and seas \citep{Sotin2012}, and bright evaporitic deposits \citep{MacKenzie2014} show up thanks to the cleared atmosphere and a better geometry of illumination. After the dissipation of the northern cloud, the geometry of several flybys allow the recording of specular reflection events, from which relevant information about the atmospheric transmission \citep{Barnes2013} or possible surface waves \citep{Barnes2014} can be retrieved.
A striking example of these reflections is visible on the T104 flyby (August 2014) in \figref{npole_cloud}, just over Kraken Mare. Finally, sporadic methane cloud activity, probably convective in nature (at the difference of the vast polar cloud described earlier) and symptomatic of the start of summer cloud activity at the northern polar region, then starts again to show up in the last series of flybys, as seen for example on the last targeted T126 flyby (April 2017) showing well defined elongated clouds between \ang{52} and \ang{58}N.

\afterpage{%
    \begin{figure*}[!ht]
    \includegraphics[width=.9\linewidth]{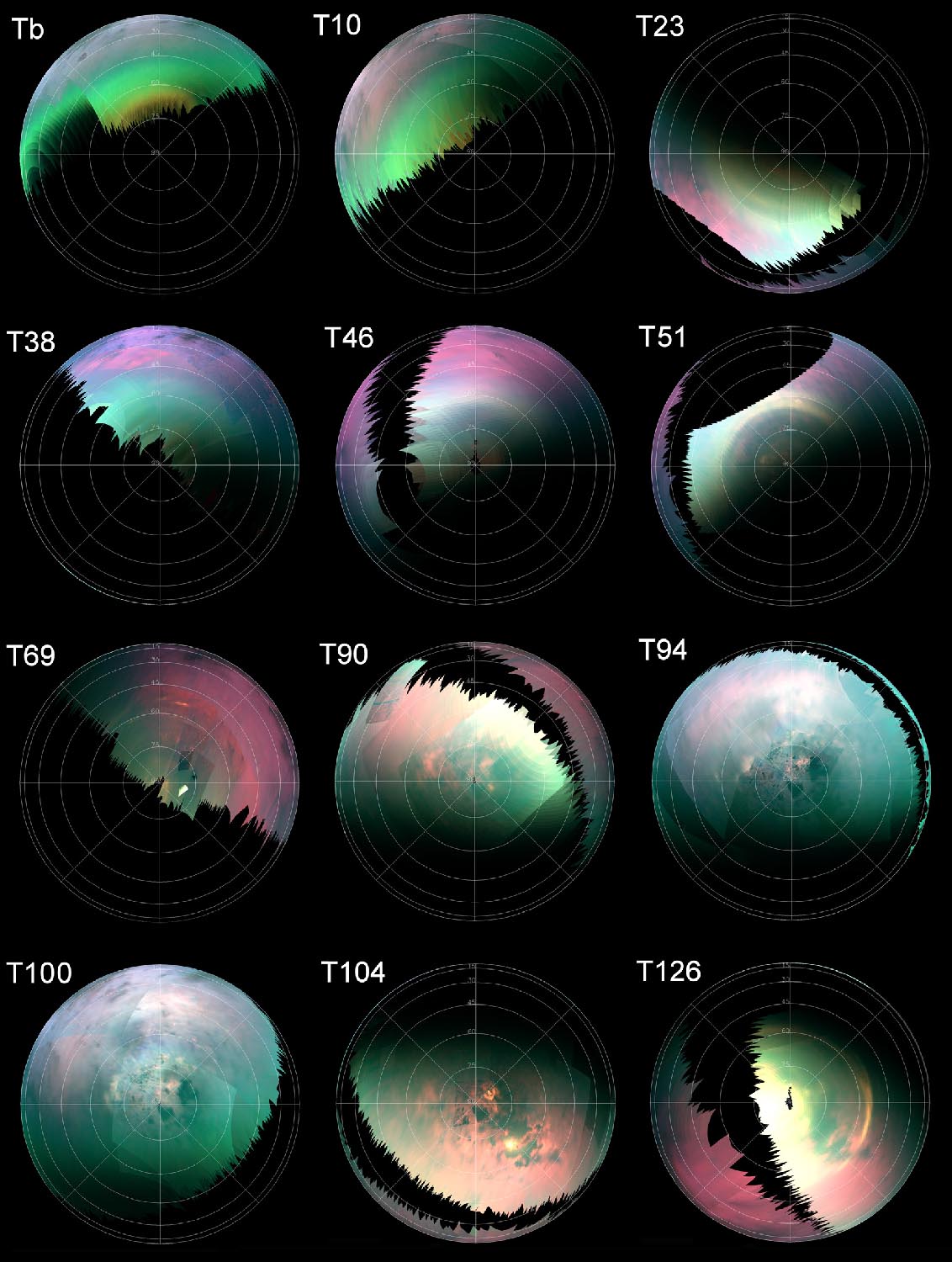}
    \caption{Evolution of the cloud and haze using orthographic projections centered on the north pole. The dark spots seen in the T94 mosaic correspond to lakes and seas that appear after the dissipation of the north polar cloud and diminishing of the polar haze opacity. Images correspond to RGB composites with red = \SI{5}{\um}, green = \SI{2.78}{\um} and blue = \SI{2.0}{\um}.}
    \label{fig:npole_cloud}
    \end{figure*}
    \clearpage
}

\section{Temporal evolution over the south pole}
Similarly to the survey performed on the north, \figref{spole_all} displays the full set of re-projected orthographic mosaics from T0 to T126 over the south pole.

\figref{spole_cloud} presents a selection of mosaics over the south pole, taken at the same time intervals as the ones shown for the north in \figref{npole_cloud} (with the exception of T100 which was replaced by T101), and with the same RGB color composite to allow inter-comparison with the \figref{npole_cloud}.
The south pole was fully illuminated at the beginning of the Cassini mission in 2004 (southern summer). Major cloud events in this area had been previously reported from Earthbased observation campaigns since December 2001 \citep{Brown2002,Roe2002,Schaller2006a}, using the adaptive optics (AO) system at the Gemini and Keck observatories.
At the time of Cassini arrival, outbursts of summer-type convective methane clouds were regularly observed over the pole during the early flybys \citep{Rodriguez2009,Rodriguez2011,Turtle2009,Turtle2011}, as shown for example on the Tb image in \figref{spole_cloud}.
A detailed detection of each individual cloud outburst can be found in \cite{Rodriguez2009, Rodriguez2011}. Small localized outbursts were consistently detected by VIMS up to December 2005, but these events appeared progressively dimmer and fainter. No clouds were observed between January (T10) and September 2006 (T18). The T23 mosaic in \figref{spole_cloud} shows a particularly clear area with no particular feature. Except for some rare events in 2008, the convective methane cloud activity over the south pole therefore ceased two years before the equinox \citep{Rodriguez2011,Turtle2009,Turtle2011}.
The limit of a diffuse but more opaque haze layer right over the pole starts to be seen progressively in T46 (11 March 2008) and T51 (27 March 2009), and more significantly on the T69 (5 June 2010) mosaic. It should be noted that VIMS targeted observations at very high spatial resolutions during the closest approach phase at T38 and T51 allowed to observe Ontario Lacus through the still optically thin haze \citep{Barnes2009,Cornet2012a,Cornet2012b}.

A major change over the south pole is observed with the onset of a polar cloud that can be seen for example in the T90 and T94 VIMS mosaics in \figref{spole_cloud}. We also provide in \figref{scloud_cubes} and \tabref{spole_cubes} a selection of the best un-projected VIMS single cube views of this cloud, from its first detection at T83 and T84 \citep[the two flybys used in][]{deKok2014} up to the last targeted T126 flyby.

This figure can serve as a catalog for those interested in going deeper into the spectral analysis of this feature with VIMS on a cube by cube basis. \tabref{spole_cubes} shows the various differences in phase angle among the numerous observations. We see for example a constant increase in the phase angle from T92 (\ang{87}) up to T99 (\ang{141}), which corresponds to less and less favorable observing conditions of the south cloud. Starting at T101 and up to T126, the phase angle stays within the range \ang{25}-\ang{85}, which results in more consistent views in the orthographic maps (see \figref{spole_all}). These different observing conditions could influence the respective amount of forward and backward scattering seen in the images.

\afterpage{%
    \begin{figure*}[!ht]
    \includegraphics[width=.9\linewidth]{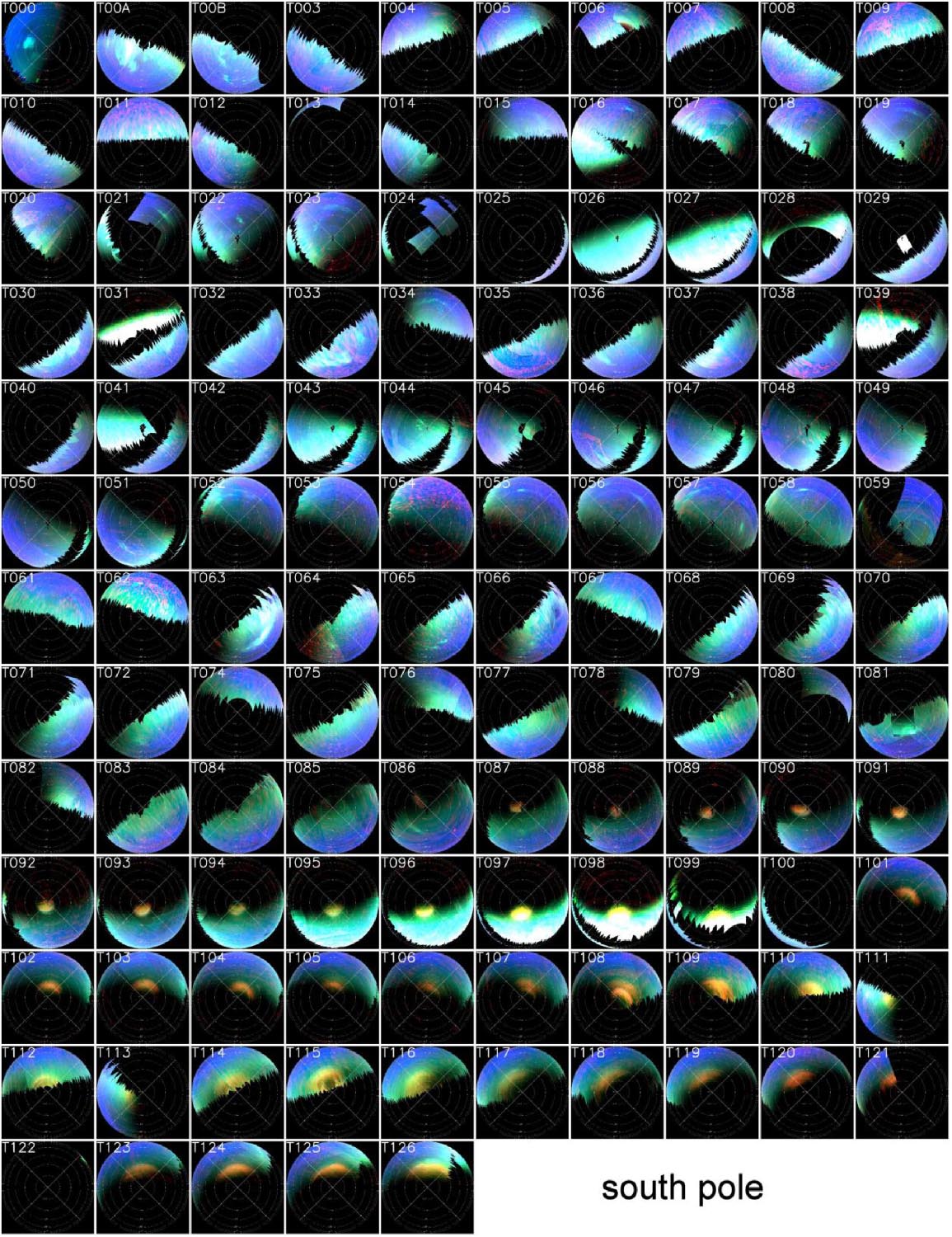}
    \caption{Systematic mapping of the south pole using all targeted Titan flybys from T0 (July 2004) to T126 (April 2017). Colors have been coded using the \SI{4.78}{\um}, \SI{2.89}{\um} and \SI{2.01}{\um} channels for the red, green and blue, respectively.}
    \label{fig:spole_all}
    \end{figure*}
    \clearpage

    \begin{figure*}[!ht]
    \includegraphics[width=.9\linewidth]{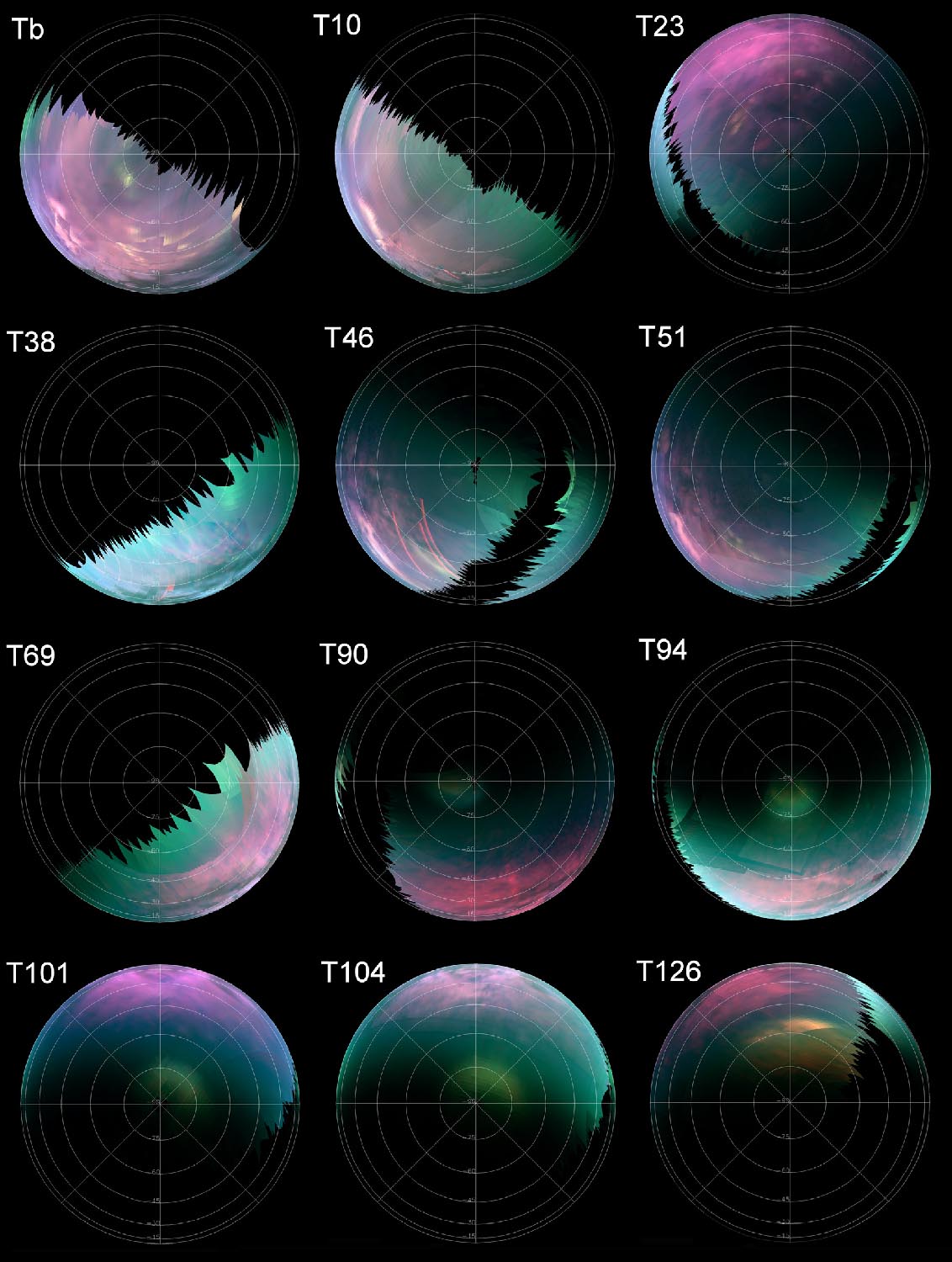}
    \caption{Views of the south pole computed on the same flybys as in \figref{npole_cloud}, using the same color scheme and color stretch for comparison. The onset of the south polar cloud started at T83 (see also \figref{scloud_cubes}). It was then observed in all subsequent flybys.}
    \label{fig:spole_cloud}
    \end{figure*}
    \clearpage

    \begin{figure*}[!ht]
    \includegraphics[width=.9\linewidth]{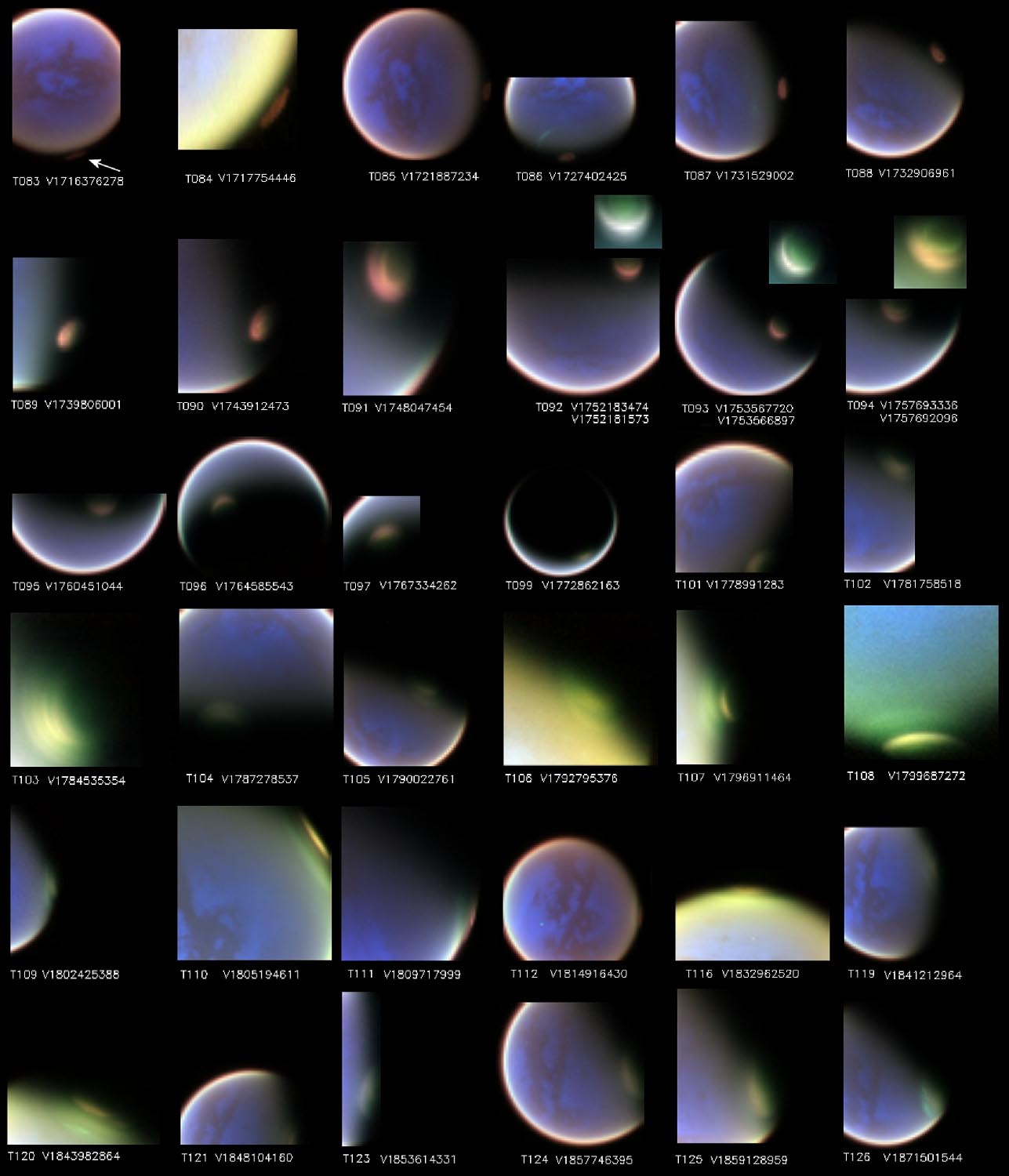}
    \caption{Selection of the most detailed VIMS individual cubes showing the southern cloud. Images are shown in their raw geometry of acquisition. Titan flyby numbers and cube labels are indicated to facilitate their identification for future dedicated studies.}
    \label{fig:scloud_cubes}
    \end{figure*}
    \clearpage
}

The very first occurrence of the southern cloud right over the limb is observed at T83 (22 May 2012, see arrow in the upper left view in \figref{scloud_cubes}), corresponding to the period where ISS also discovered the feature using high spatial resolution images \citep{West2016}. It is then consistently seen at each flyby, as shown by the
reddish tones in the mosaics in \figref{spole_all}, the selection of maps reproduced in \figref{spole_cloud} and raw unprojected views (\figref{scloud_cubes}).
A simple geometric consideration using the surface curvature at T83 and T84 indicates that the observed cloud over the south pole is at an altitude of \SI{300(70)}{km} above the surface. This is consistent with the \SI{300(10)}{km} found by \cite{West2016} using ISS images. We also note that month after month, the spatial coverage of the cloud increases. This is illustrated in \figref{scloud_closeup}, which displays a selection of zooms over the south pole. We see a latitudinal limit starting at \ang{75}S in the early stages in 2013, and extending progressively to \ang{68}S in 2014, to \ang{60}S mid-June 2016, to finally reach \ang{58}S in April 2017. This limit is close to the one observed on the north polar cloud at the beginning of the Cassini mission.

\afterpage{%
    \begin{figure}[!ht]
    \vspace{2.75cm}
    \includegraphics[width=.95\linewidth]{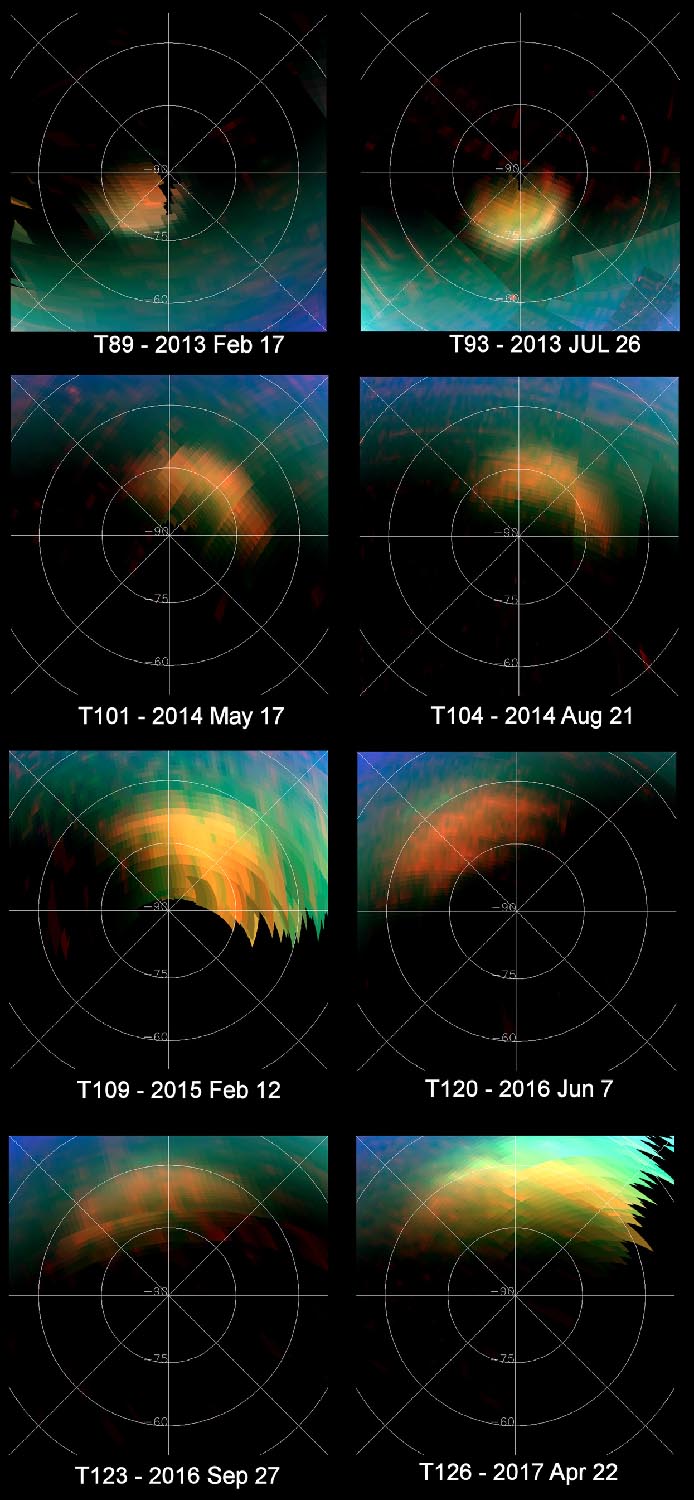}
    \caption{Evolution of the cloud over the south pole. The extent in latitude ranges from \ang{75}S in the first stages in 2013 up to \ang{58}S near the end of the mission in 2017. Colors correspond to R = \SI{4.78}{\um}, G = \SI{2.78}{\um}, B = \SI{2.01}{\um}. A reddish tone indicates a potential HCN component.}
    \label{fig:scloud_closeup}
    \end{figure}

    \begin{table}[!ht]
    \caption{List of the most detailed VIMS single cubes of the southern polar cloud, shown in \figref{scloud_cubes}.}
    \label{tab:spole_cubes}
    \begin{tabular}{l l l l}
    \toprule
    Flyby  & Date & Cube ID & Phase angle \\
    \midrule
    T83 & 05/22/2012 & CM\_1716376278   & \ang{22} \\
    T84 & 06/06/2012 & CM\_1717754446   & \ang{27} \\
    T85 & 07/24/2012 & CM\_1721887234   & \ang{36} \\
    T86 & 09/26/2012 & CM\_1727402425   & \ang{46} \\
    T87 & 11/13/2012 & CM\_1731529002   & \ang{52} \\
    T88 & 11/29/2012 & CM\_1732906961   & \ang{60} \\
    T89 & 02/17/2013 & CM\_1739806001   & \ang{67} \\
    T90 & 04/05/2013 & CM\_1743912473   & \ang{72} \\
    T91 & 05/23/2013 & CM\_1748047454   & \ang{78} \\
    T92 & 07/10/2013 & CM\_1752183474   & \ang{87} \\
        &            & CM\_1752181573   & \ang{87} \\
    T93 & 07/26/2013 & CM\_1753567720   & \ang{91} \\
        &            & CM\_1753566897   & \ang{92} \\
    T94 & 09/12/2013 & CM\_1757693336   & \ang{100} \\
        &            & CM\_1757692096   & \ang{100} \\
    T95 & 10/14/2013 & CM\_1760451044   & \ang{109} \\
    T96 & 12/01/2013 & CM\_1764585543   & \ang{116} \\
    T97 & 01/01/2014 & CM\_1767334262   & \ang{124} \\
    T99 & 03/06/2014 & CM\_1772862163   & \ang{141} \\
    T101    & 05/17/2014 & CM\_1778991283   & \ang{66} \\
    T102    & 06/18/2014 & CM\_1781758518   & \ang{74} \\
    T103    & 07/20/2014 & CM\_1784535354   & \ang{85} \\
    T104    & 08/21/2014 & CM\_1787278537   & \ang{82} \\
    T105    & 09/22/2014 & CM\_1790022761   & \ang{73} \\
    T106    & 10/24/2014 & CM\_1792795376   & \ang{64} \\
    T107    & 12/10/2014 & CM\_1796911464   & \ang{58} \\
    T108    & 01/11/2015 & CM\_1799687272   & \ang{50} \\
    T109    & 02/12/2015 & CM\_1802425388   & \ang{44} \\
    T110    & 03/16/2015 & CM\_1805194611   & \ang{37} \\
    T111    & 05/07/2015 & CM\_1809717999   & \ang{67} \\
    T112    & 07/07/2015 & CM\_1814916430   & \ang{25} \\
    T116    & 02/01/2016 & CM\_1832962520   & \ang{29} \\
    T119    & 05/06/2016 & CM\_1841212964   & \ang{53} \\
    T120    & 06/07/2016 & CM\_1843982864   & \ang{61} \\
    T121    & 07/25/2016 & CM\_1848104160   & \ang{66} \\
    T123    & 09/27/2016 & CM\_1853614331   & \ang{68} \\
    T124    & 11/14/2016 & CM\_1857746395   & \ang{68} \\
    T125    & 11/29/2016 & CM\_1859128959   & \ang{67} \\
    T126    & 04/22/2017 & CM\_1871501544   & \ang{71} \\
    \bottomrule
    \end{tabular}
    \end{table}
    \clearpage
}

\section{Temporal evolution of the HCN spectral signatures over the two poles}

The development of a cloud of HCN during northern winter has been suspected since the Voyager flyby \citep[e.g.][]{Coustenis1999,Samuelson2007,Anderson2010}.
The first mention of a detection of condensed HCN signatures with VIMS on the north polar cloud was made by \cite{Clark2010}. We have used our new set of global mosaics to refine this detection, and to systematically search for such spectral features on both poles in all targeted flybys. \figref{spectra_HCN}a presents a comparison of a typical VIMS spectrum taken within the northern cloud (red curve) and a VIMS spectrum taken outside of the cloud (black curve), during the Tb observation in December 2004. The main difference is the presence of two diagnostic peaks at \num{3.21} and \SI{4.78}{\um}, which are indicative of the presence of either solid or liquid HCN \citep{Clark2010}.

It is interesting to see that this spectral signature, which appears as an orange-reddish tint in all the mosaics shown in Figs. \autoref{fig:npole_all}, \ref{fig:spole_all} and \ref{fig:scloud_closeup} can be tracked down on the north from the beginning of the mission up to the T42-T45 flybys. The growing south polar cloud exhibits similar spectral signatures, as was first reported by \cite{deKok2014} using two flybys in May and June 2012. \figref{spectra_HCN}b shows a comparison between two VIMS spectra taken within and outside the southern vortex at T91, where the HCN diagnostic peak is also clearly detected.

\afterpage{%
    \begin{figure*}[!ht]
    \includegraphics[width=.8\linewidth]{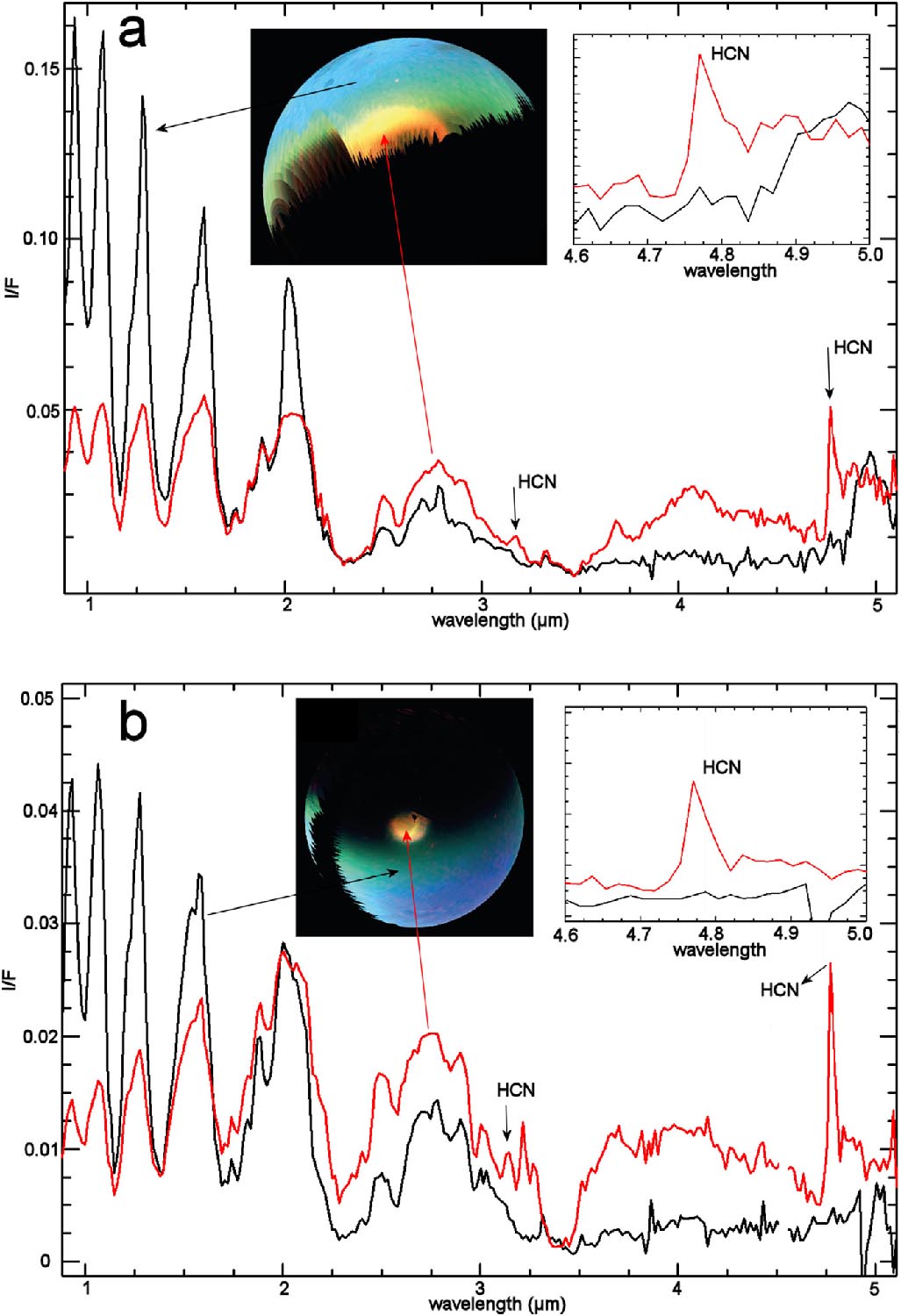}
    \caption{Comparison of VIMS spectra acquired within (red) and outside (black) the north polar cloud at Tb in December 2004 (up) and at T91 in May 2013 (bottom). One of the main differences is the presence of two diagnostic peaks at \SI{3.21}{\um} and \SI{4.78}{\um} indicating the presence of HCN.}
    \label{fig:spectra_HCN}
    \end{figure*}
    \clearpage
}

\begin{figure*}[!ht]
\includegraphics[width=.84\linewidth]{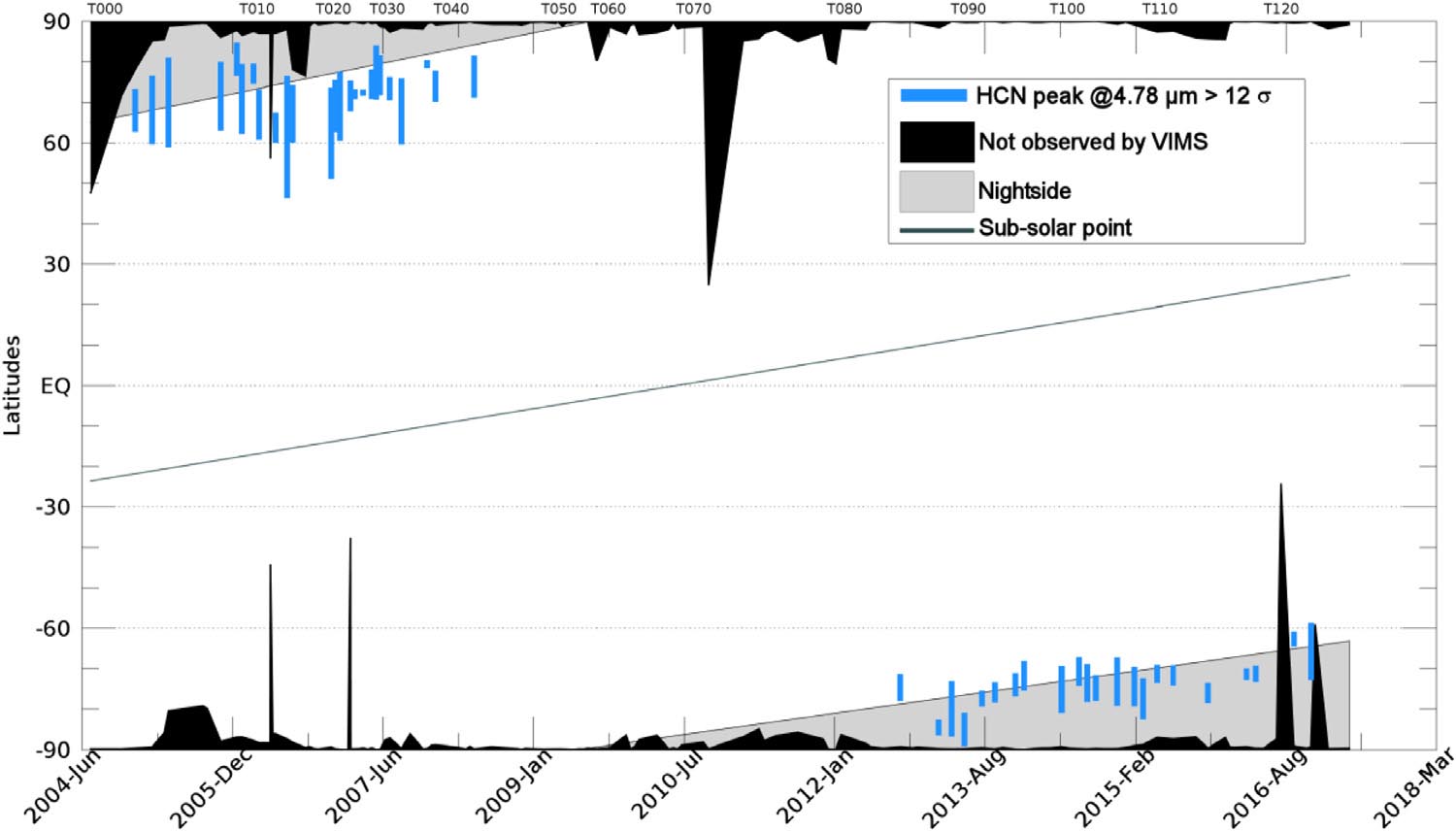}
\caption{Spectral detection (in blue) of the 4.78 μm condensed HCN peak with VIMS during all targeted Titan flybys from 2004 to 2017. HCN Spectral features are found over the north pole between 2004 and 2008, and then appear over the south pole from 2012 to 2017.}
\label{fig:cloud_coverage}
\end{figure*}

We have then used the height of the \SI{4.78}{\um} peak as a spectral criterion to systematically identify with VIMS the condensed HCN signature at both poles in all flybys. \figref{cloud_coverage} shows a summary of all the detections which have been obtained from our set of global hyperspectral mosaics. It shows the extension in latitude, averaged over all longitudes, of the areas where the \SI{4.78}{\um} peak is 12 times stronger than the typical noise (sigma), evaluated using the \num{4.0}-\SI{4.7}{\um} more neutral spectral range. We chose a threshold of 12 sigma to be very conservative, with the same value in all flybys to allow the comparison. The condensed HCN detections appear in blue in \figref{cloud_coverage}. The black zones corresponds to the areas which were not observed by VIMS. This systematic spectral study confirms the visual analysis made from the color trends in orthographic mosaics in Figs. \autoref{fig:npole_all}, \ref{fig:spole_all} and \ref{fig:scloud_closeup}, with a systematic presence of condensed HCN detected in the north from 2004 to 2008 and then an apparition of condensed HCN signatures in the south between 2012 and 2017. The general latitudinal retreat for the northern hemisphere and the extension for the southern hemisphere also appear in this HCN signature evolution. A detailed comparison with the detections made by other more sensitive instruments such as CIRS could be envisaged in further studies, but this falls beyond the scope of this paper.

The presence of HCN particles within the clouds, together with temperature determinations from mid-infrared observations, provides constrains on the cooling of Titan's atmosphere inside the winter polar vortex. \cite{deKok2014} noted how far out of thermochemical equilibrium the southern polar HCN cloud apparently was, and concluded that the post-equinox cooling at the winter pole seems to be much more efficient than previously thought. The detailed study of the thermochemical condition in the southern pole at equinox involves the interplay of several climate components: dynamics, thermo and photochemistry, microphysics and radiative transfer. Circulation was feeding the north pole with aerosols and gases before the equinox. After 2009 (equinox) the circulation started to reverse, and in about three years, it reached a state where it was feeding the south polar region. \cite{Teanby2017} evaluated the role of each component during this period and suggested that the cooling was primarily triggered by the extreme accumulation of trace gas species in the southern polar region, during the seasonal reversal that increased the atmosphere IR cooling efficiency. It occurred that this radiative cooling was even stronger than the adiabatic heating due to descending circulation and produced a rapid and strong cooling of the atmosphere. This further triggered a chain of events (condensation of species including HCN, removal of species, return to a normal situation) in a period of rapid seasonal change (polar illumination and circulation turnover). This excursion out of equilibrium was a transient event and could only concern the south pole where species and aerosols started to accumulate. On other hand, the clouds observed at the north pole come from a steady state condition (2004-2009) which gradually changed and drove to the gradual disappearance of the clouds when the north polar region was no more fed by circulation. To go one step further with VIMS into this issue of thermochemical equilibrium over the course of the Cassini mission, a more comprehensive temporal comparison with other data sets such as CIRS, which provides additional constrains on temperature profiles, would be needed, in addition to radiative transfer modeling to retrieve the altitude of the HCN cloud over time. This comparison is left for further studies.

\section{Conclusion}

\begin{figure*}[!ht]
\includegraphics[width=.95\linewidth]{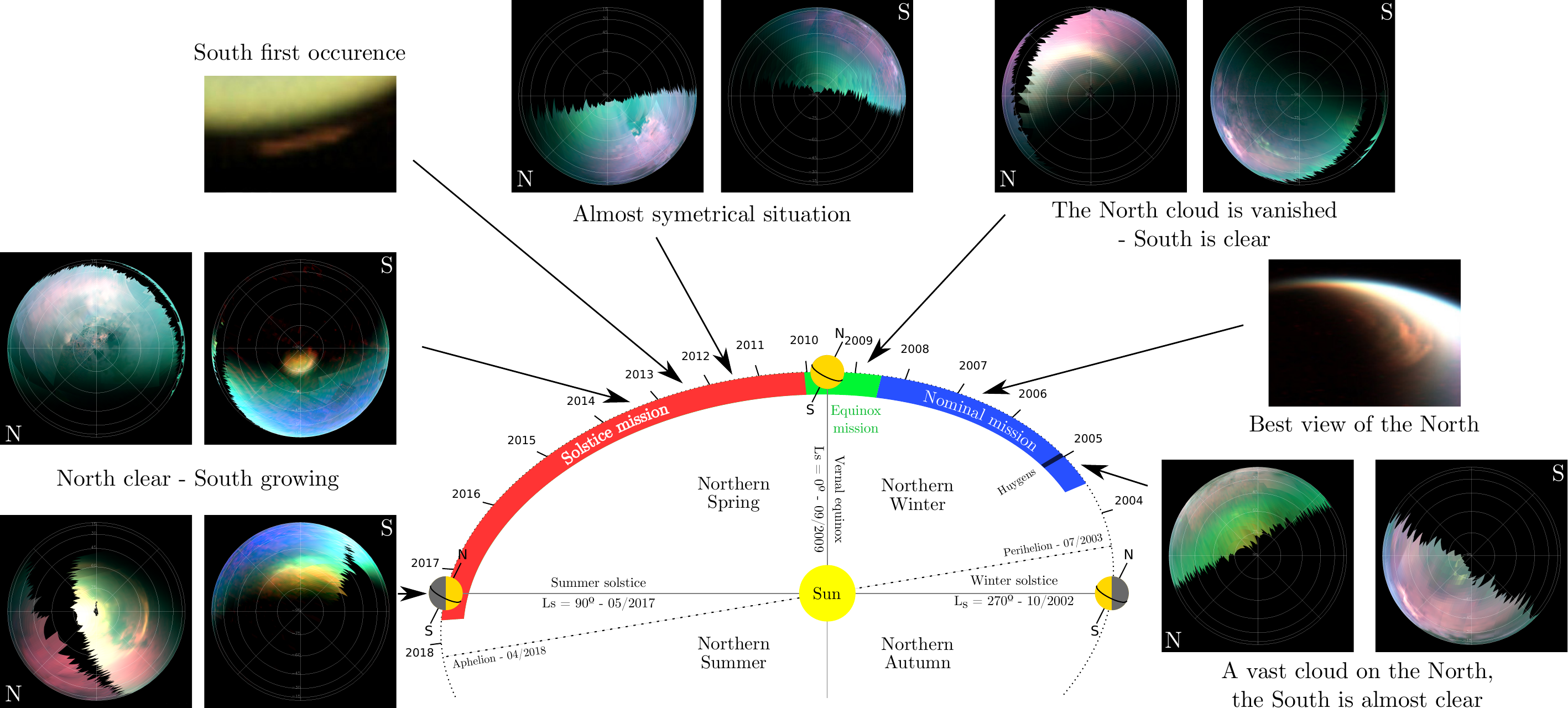}
\caption{Summary of the cloud mapping at both poles during the Cassini mission.}
\label{fig:orbit}
\end{figure*}

We have performed a complete analysis of the VIMS global archive to compute hyperspectral maps for each targeted flyby of Titan from T0 in July 2004 to T126 in April 2017. Using false color composites, we were able to illustrate the temporal changes which occurred at both poles of Titan during the entire Cassini mission. A summary of the observations is provided in \figref{orbit}.

Global changes in polar cloud coverage have been observed in Titan's atmosphere and stratosphere. They are linked to the reversal of the main circulation cell which occurred around the equinox in 2009. With the complete disappearance of the north polar cloud after 2009, which is consistent with the seasonal change of circulation predicted by the GCM used in \cite{Rannou2006}, one of the most striking seasonal changes observed is the progressive
onset of a polar cloud over the south pole, which started to appear in VIMS images in May 2012. We were able to detect a consistent condensed HCN spectral signature over the north pole in most of the flybys from 2004 to 2008, and then at the south pole in flybys from 2012 to 2017. Provided that the north and south behave symmetrically with time, the situation we have been witnessing in the south since 2012 occurred around 1998 in the north, when Cassini just started its seven years cruise to Saturn, and will occur again around 2028.
The evolution at the south pole might therefore represent an early stage of what has been observed by Cassini in the north during the first Titan flybys. We note that no cloud at \SI{300}{km} was detected in the northern polar region. The HCN in the north should be located in a cloud at lower altitude before 2009. It could be part of the cloud at \SI{60}{km} observed by \cite{Griffith2006,LeMouelic2012}.
It should also be noted that another thin layer of scatterers exists above this main cloud, around \SI{100(20)}{km}, as shown by \cite{Rannou2012}. In any case, we may conclude that the HCN south polar cloud observed at \SI{300}{km}, in 2012, and the HCN in a putative north pole cloud at lower altitude before 2009 are the end members of the HCN cloud cycle. Further studies using detailed radiative transfer and photometric models would be needed in order to quantitatively characterize the altitude where HCN particles are located, together with their amount and size. This would allow to capture the subtle differences and similarities between the two polar clouds. Such information would be crucial to better understand Titan's climate and would be valuable to constrain Global Climate Model with cloud microphysics. Cassini has been able to follow the seasonal changes during almost half a Titan year. After the end of Cassini in September 2017, some limited monitoring might still be possible using the forthcoming James Webb Space Telescope (JWST) and the European Extremely Large Telescope (E-ELT), waiting for the arrival of a new dedicated spacecraft.

\section*{Acknowledgment}
Authors are very grateful to R. West and an anonymous reviewer for their insightful reviews, which greatly improved this paper. This work has been partly funded by the French spatial agency (CNES) and the Institut Universitaire de France. We also acknowledge financial support from the UnivEarthS LabEx program of Sorbonne Paris Cite (\texttt{ANR-10-LABX-0023} and \texttt{ANR-11-IDEX-0005-02}) and the French National Research Agency (\texttt{ANR-APOSTIC-11-BS56-002} and \texttt{ANR-12-BS05-001-3/EXO-DUNES})

\vfill\null
\clearpage
\bibliography{Biblio}

\end{document}